\documentclass[a4paper,11pt]{article}
\pdfoutput=1

\usepackage{jcappub}
\usepackage[T1]{fontenc}
\usepackage[utf8]{inputenc}
\usepackage{lmodern}
\usepackage{multirow} 
\usepackage{subcaption} 
\usepackage{float} 

\usepackage{xurl} 
\usepackage{hyperref}
\DeclareRobustCommand{\mytt}[1]{\ifmmode\text{\path{#1}}\else\path{#1}\fi}

\usepackage{booktabs}
\usepackage{xcolor}

\definecolor{valecol}{rgb}{0,0.5, 1.}

\definecolor{rmcol}{rgb}{1.,0.5, 0.}
\definecolor{rmtol}{rgb}{1,0,0}

\graphicspath{{./figs/}}

\def\d{{\rm d}}

\newcommand{\Om}{\Omega_\mathrm{m}}

\newcommand{\Ok}{\Omega_\mathrm{k}}

\arxivnumber{2510.18752}

\title{\boldmath Probing cosmology with bright sirens from the CosmoDC2\_BCO LSST synthetic catalog}

\author[a,b]{Ranier Menote}
\author[b,c,d,e]{and Valerio Marra}

\affiliation[a]{PPGCosmo, Universidade Federal do Espírito Santo, 29075-910, Vitória, ES, Brazil}
\affiliation[b]{Laboratório Interinstitucional de e-Astronomia, 20921-400, Rio de Janeiro, RJ, Brazil}
\affiliation[c]{Dep.\ de Física, Universidade Federal do Espírito Santo, 29075-910, Vitória, ES, Brazil}
\affiliation[d]{INAF -- Osservatorio Astronomico di Trieste, 34131 Trieste, Italy}
\affiliation[e]{IFPU -- Institute for Fundamental Physics of the Universe, 34151, Trieste, Italy }

\emailAdd{ranier\_m@hotmail.com}
\emailAdd{valerio.marra@me.com}

\abstract{
Bright sirens, i.e.\ gravitational-wave detections of compact binary mergers with electromagnetic counterparts, provide a self-calibrated distance–redshift relation and are therefore powerful probes of cosmic expansion. Using the \texttt{CosmoDC2\_BCO} catalog, we forecast cosmological constraints from current (LVK) and next-generation (ET, CE) detector networks, in combination with a Roman-like Type Ia supernova sample. We find that third-generation networks reach sub-percent precision on the Hubble constant within a few years, achieving 0.2\% after a decade with CE+ET+LVK, while LVK remains limited to the 6\% level.
The LVK fifth observing run may shed light on the $H_0$ tension \textit{only if} the inferred value falls outside the range spanned by the Planck and SH0ES determinations, which currently achieve far higher precisions.
Supernovae do not directly tighten \(H_0\) but stabilize its inference through parameter correlations and enable an absolute calibration of the supernova magnitude \(M_B\). In dynamical dark-energy models, the joint analysis of Roman supernovae and bright sirens yields a Figure of Merit of 25 for ET+LVK and 76 for CE+ET+LVK, to be compared with the state-of-the-art DESI DR2 BAO plus DESY5 supernovae value of 56. Sky-localization thresholds of \(\Delta\Omega<50~\mathrm{deg}^2\), or even \(\Delta\Omega<10~\mathrm{deg}^2\), entail only mild penalties, suggesting efficient follow-up strategies. These results establish third-generation GW+EM observations, especially when combined with Roman supernovae, as a cornerstone for precision cosmology in the next decade.
}

\keywords{neutron stars, kilonovae, gravitation waves, cosmology, dark energy}

\begin{document}
\maketitle
\flushbottom

\section{Introduction}
\label{sec:intro}

The detection of gravitational waves (GWs) from compact binary mergers has opened a new avenue for precision cosmology \citep{LIGOScientific:2025jau}. When a GW event is accompanied by an electromagnetic (EM) counterpart—most notably a kilonova associated with a binary neutron star (BNS) or neutron star–black hole (BHNS) merger—the source redshift can be identified independently of the GW signal. These `bright sirens' provide a self-calibrated distance–redshift relation that is anchored in fundamental physics and does not rely on the traditional distance ladder. As such, they offer a powerful and conceptually clean route to measure the Hubble constant, \(H_0\), and to test the expansion history at low and intermediate redshift.
A milestone in this context was GW170817, the first BNS merger detected in gravitational waves \citep{LIGOScientific:2017vwq}, which marked the beginning of multi-messenger astronomy \citep{LIGOScientific:2017ync}. It was accompanied by a short gamma-ray burst and a kilonova—a faint, rapidly evolving transient powered by radioactive decay—and provided the first direct demonstration of the bright-siren method \citep{Abbott:2017xzu,Palmese:2023beh}.

Motivated by the ongoing debate over the value of \(H_0\) \citep{Abdalla:2022yfr,CosmoVerseNetwork:2025alb} -- with the Planck CMB observations and the local Cepheid–SNe~Ia calibration yielding discrepant results at the \(\sim 5\sigma\) level \citep{Planck:2018vyg,Riess:2021jrx} -- we investigate the constraining power of bright sirens detected by current and next-generation GW networks, both on their own and in combination with a Roman-quality SNe~Ia dataset. While bright sirens are exceptionally sensitive to \(H_0\), supernovae provide complementary leverage on the matter density and the dark-energy equation of state, and, when combined with bright sirens, allow an absolute calibration of the SNe luminosity scale. This synergy represents a key objective for the coming decade, as electromagnetic facilities such as the Rubin Observatory and the Nancy Grace Roman Space Telescope undertake wide and deep time-domain surveys that will enable the systematic discovery and characterization of kilonovae and Type~Ia supernovae. 

To quantify these prospects, we build on the \texttt{CosmoDC2\_BCO} synthetic catalog \citep{Menote:2025zmn}, which embeds binary compact object mergers within the \texttt{CosmoDC2} galaxy catalog \citep{LSSTDarkEnergyScience:2019hkz} and assigns EM counterparts where appropriate. We consider three detector configurations: the current LIGO–Virgo–KAGRA (LVK) network, the Einstein Telescope \citep[ET,][]{ET:2019dnz} in combination with LVK, and the third-generation Cosmic Explorer \citep[CE,][]{Reitze:2019iox} plus ET plus LVK. We model detectability and parameter recovery for GW events using waveform families from \texttt{IMRPhenomX}, apply realistic duty cycles, and impose sky-localization cuts representative of Target-of-Opportunity (ToO) follow-up. For SNe~Ia, we adopt a Roman-like redshift distribution and per-bin effective uncertainties that include intrinsic dispersion and lensing scatter. We use Fisher forecasts for large samples and dedicated MCMC analyses where the Gaussian approximation may break down (notably for LVK events).

We explore three cosmological models -- \(\Lambda\)CDM,  \(w\)CDM, and  \(w_0w_a\)CDM -- along with a low-redshift, model-independent cosmographic analysis of \(H_0\), analogous to the SH0ES analysis. The primary goals are to: (i) chart the time evolution of the \(H_0\) precision attainable with bright sirens across detector generations; (ii) assess how the joint GW+SNe analysis breaks degeneracies and calibrates the SNe absolute magnitude; (iii) quantify the sensitivity to dark-energy evolution via the Figure of Merit in the \(w_0\)–\(w_a\) plane; and (iv) evaluate the impact of sky-localization thresholds that bracket realistic EM follow-up strategies.

This paper is organized as follows. In Section~\ref{sec:data} we describe the synthetic datasets employed, including the GW catalog and kilonova selection, and the Roman supernova sample. Section~\ref{Methodology} outlines our methodology, detailing the cosmological models, the time-evolution strategy, and the likelihood construction and priors. Our main parameter constraints and comparisons across detector networks and data combinations are presented in Section~\ref{sec:results}. We conclude with a summary and outlook in the Section~\ref{sec:conclusions}.

\section{Synthetic data}
\label{sec:data}

The fiducial cosmology underlying the synthetic data used in this work is the same as that adopted by the \texttt{CosmoDC2\_BCO}~\citep{Menote:2025zmn} and \texttt{CosmoDC2}~\citep{LSSTDarkEnergyScience:2019hkz} catalogs, namely a flat \(\Lambda\)CDM model with parameter values \(\{H_0, \Om, w_0, w_a, \Ok\} = \{71,\, 0.2648,\, -1,\, 0,\, 0\}\). In terms of the cosmographic expansion, this corresponds to fiducial values of \(q_0 = -0.6028\) and \(j_0 = 1\).

\subsection{GW data}


Here we make use of the \texttt{CosmoDC2\_BCO} synthetic catalog of gravitational-wave events with electromagnetic counterparts~\citep{Menote:2025zmn}, publicly available at \url{https:/github.com/LSSTDESC/CosmoDC2_BCO} \citep[][Version 1]{menote2025-zenodo}.  
This dataset was constructed by embedding binary compact object (BCO) mergers into the \texttt{CosmoDC2} LSST synthetic galaxy catalog~\citep{LSSTDarkEnergyScience:2019hkz}, assigning electromagnetic counterparts where relevant. For each detected GW event, the pipeline provides forecasted covariance matrices for the inferred source parameters.
A comprehensive description of both catalogs is provided in~\citep{LSSTDarkEnergyScience:2019hkz,Menote:2025zmn}; here and in the following section we restrict ourselves to a brief overview.

The merger rate of compact binaries was derived from population-synthesis simulations~\citep{Santoliquido:2022kyu} and mapped onto the galaxy distribution of \texttt{CosmoDC2} using a combination of empirical relations and semi-analytic modeling.  
Events were then sampled in bins of redshift and stellar mass, with each selected galaxy hosting one merger.  
The resulting GW dataset corresponds to a 10-year observation period and, to emulate the LSST survey footprint, was constructed from 36 statistically equivalent sky tiles.  
Each tile covers about 440~deg$^2$, together spanning nearly the full southern sky for a total effective area of approximately 16\,000~deg$^2$.

Binary properties were assigned following observationally motivated distributions~\citep{KAGRA:2021duu,Iacovelli:2022bbs}. 
Black-hole binaries adopted mass and spin distributions calibrated from LVK detections, while neutron stars were restricted to narrower mass ranges and low spin values.  
Spin orientations were sampled from a mixture of isotropic and aligned components.  
Extrinsic parameters, including orientation angles and sky positions, were taken directly from the corresponding host galaxies. For instance, neutron stars follow a mass distribution modeled by an edge-smoothed power law in the range $1.2$--$2.0\,M_\odot$, while black holes in BHNS systems are drawn from a uniform mass distribution spanning $3$--$60\,M_\odot$. All these population distributions are fully described and explored in the \texttt{CosmoDC2\_BCO} catalog release paper \cite{Menote:2025zmn}.

Waveforms from the \texttt{IMRPhenomX} family were employed, including variants with higher-order multipoles and spin precession.  
For the systems of primary interest in this work, BNS and BHNS, the adopted models were \texttt{IMRPhenomXHM} and \texttt{IMRPhenomXPHM}, respectively. The \texttt{XHM} model improves the description of the signal by incorporating subdominant multipoles, while \texttt{XPHM} extends this by also accounting for spin precession effects.
Both types of systems are in principle affected by tidal deformability, parametrized by the quantities $\Lambda_i$, which modify the phase evolution of the waveform.  
These corrections enter at higher post-Newtonian order and are suppressed by factors proportional to $q^4$~\citep{Vines:2011ud}, where $q$ is the mass ratio.  
As a result, tidal effects are negligible for BHNS systems, particularly in the regime where the black hole is much more massive than the neutron star. For BNS systems, however, tidal interactions can significantly impact the late inspiral. Incorporating such corrections requires modeling of the neutron star equation of state, which introduces additional theoretical uncertainties. Furthermore, tidal effects are not yet implemented in the current version of the \texttt{GWDALI} code used for waveform generation and inference~\citep{deSouza:2023ozp}.
Since tidal parameters are largely uncorrelated with luminosity distance at the signal-to-noise ratios considered here, omitting tidal corrections leads only to subdominant biases in the cosmological inference. Likewise, spin precession is not expected to play a significant role for BNS systems, given their typically low spins. For these reasons, the BNS dataset was generated using the \texttt{XHM} waveform model, which provides a suitable balance between realism and computational efficiency for our cosmological forecasts.

Events were considered detected if their signal-to-noise ratio exceeded the network threshold, $\rho_{\rm net} > 12$, where $\rho_{\rm net} = \sqrt{\sum_i \rho_i^2}$ and $\rho_i$ denotes the individual optimal (matched-filter) SNR:
\begin{equation}
\rho_i^2 = 4 \int_0^\infty 
\frac{|h_i(f)|^2}{S_{n,i}(f)}\, \d f \,,
\end{equation}
where $h_i(f)$ is the frequency-domain strain response of that detector and $S_{n,i}(f)$ is its one-sided noise power spectral density. 
Parameter uncertainties were estimated with the Fisher-matrix approximation implemented in the \texttt{GWDALI} package, and validated against selected full MCMC analyses.  
A complete Bayesian parameter estimation of a gravitational-wave signal requires sampling over 15 parameters, describing both the intrinsic properties of the binary (masses and spins) and the extrinsic properties (orientation, distance, and sky position).  
While feasible for a small number of events, such analyses become computationally prohibitive for catalogs containing thousands of mergers.

The \texttt{CosmoDC2\_BCO} catalog provides a collection of simulated gravitational-wave events, each characterized through the computation of a Fisher information matrix associated with the likelihood of the GW signal $h(\vec{\theta}_{\rm gw})$ detected by a given detector network.
Here $\vec{\theta}_{\rm gw}$ denotes the set of gravitational–wave parameters (e.g.\ $d_L$, $\mathrm{RA}$, $\mathrm{Dec}$, and the remaining intrinsic and extrinsic parameters).
In this framework, the catalog used as the data source for this work consists of a set of Fisher matrices of the form:
\begin{equation}
\Gamma^{(k)}_{ab}
= \left(
\frac{\partial h^{(k)}}{\partial \theta_a}
\,\Bigm|\,
\frac{\partial h^{(k)}}{\partial \theta_b}
\right),
\end{equation}
where $(a|b)$ denotes the standard noise-weighted inner product, and the index $k$ labels individual detectors in the network. The corresponding network Fisher matrix is obtained by summing over detectors,
\begin{equation}
\Gamma^{\rm (net)}_{ab} = \sum_k \Gamma^{(k)}_{ab}.
\label{fisher_net}
\end{equation}
From the inverse of the network Fisher matrix, the covariance matrix is obtained, providing access to the marginalized uncertainties of the GW parameters. In particular, the uncertainty on the luminosity distance plays a central role in the cosmological analyses presented in this work, as clearly pointed in the Section~\ref{sec:like}.

In this work, we also employ sky-localization information to filter out BCO events with poor localization.  
This is quantified by the error-ellipse area $\Delta \Omega$, defined as
\begin{equation}
\Delta\Omega_X
= -\,\Delta\Omega_s\,
\ln\!\left(1 - \frac{X}{100}\right),
\end{equation}
where $X$ denotes the confidence level. The characteristic angular scale is given by
\begin{equation}
\Delta\Omega_s
= 2\pi\,|\sin\delta|\,
\sqrt{\det \mathcal{C}_{\delta\phi}},
\end{equation}
with $\mathcal{C}_{\delta\phi}$ being the covariance submatrix in right ascension and declination, and the factor $|\sin\delta|$ accounting for the spherical-coordinate projection. See \cite{Menote:2025zmn} for more details. 
Only events with sufficiently small $\Delta \Omega$ are retained for the cosmological analysis.


Events were propagated through different gravitational-wave detector networks, with realistic duty cycles applied.  
In this work we consider three representative setups, each operating with a 70\% duty cycle:
\begin{itemize}

\item \textbf{LVK:} The current second-generation network of LIGO, Virgo, and KAGRA, taken at their projected design sensitivities for the upcoming O5 observing run. This setup represents the near-future baseline capability of the gravitational-wave community. \footnote{LIGO--India is not yet included in the current \texttt{CosmoDC2\_BCO} catalogs, although its implementation is planned for future releases.}

\item \textbf{ET+LVK:} An extended configuration that combines the full LVK network with the next-generation Einstein Telescope (ET) in its triangular configuration, located in the Euregio Meuse-Rhine region \cite{Menote:2025zmn}, providing higher sensitivity and redshift reach.

\item \textbf{CE+ET+LVK:} A third-generation network including both the Einstein Telescope and the Cosmic Explorer (CE), located near Livingston (USA) \cite{Menote:2025zmn}, in addition to the LVK detectors, representing the most powerful configuration considered here.

\end{itemize}

It is worth emphasizing that there is no guaranteed operational plan ensuring that the current 2G LVK detectors will overlap with 3G facilities such as ET or CE. The detector combinations explored in this work should therefore be interpreted as forecast scenarios rather than firm predictions of future observing timelines. Their purpose is to quantify the scientific return under plausible network assumptions and to illustrate how the cosmological performance depends on the adopted configuration.


Although 3G detectors will be far more sensitive than 2G instruments, recent studies \cite{Nitz:2021pbr,Gupta_2024,Menote:2025zmn} have stressed that sensitivity alone does not guarantee robust sky localization and three-dimensional event reconstruction when the network has only one or two sites. These capabilities are critical for identifying bright sirens and for most cosmological applications. For example, Ref.~\cite[][Fig.~12]{Menote:2025zmn} shows that the 2ET-L configuration \footnote{This configuration, alternative to the triangular ET design, consists of two detectors with the same individual sensitivity and dimensions as the standard ET, but arranged in an L-shaped configuration and located in Sardinia and the Euregio Meuse-Rhine region. The triangular configuration alone is not considered here, as it provides comparatively poor source localization.} 
yields substantially larger sky-localization error ellipses, $\Delta\Omega$, than the extended 2ET-L+LVK network. Likewise, adding LVK to a two-site 3G network (e.g.\ CE+ET) markedly improves the localization of BNS events, whereas the contribution of LVK becomes marginal only once three 3G detectors operate simultaneously. While a three-detector 3G network would largely mitigate these limitations, it is arguably a more speculative assumption than continued 2G+3G overlap, which is also considered in other analyses \cite{Loffredo:2024gmx}.


An intermediate possibility is to rely on LIGO-India, which is currently expected to begin operations in the early 2030s and could overlap with the early 3G era \cite{ET:2025xjr}. However, this schedule remains subject to construction and commissioning uncertainties. In this context, assuming continued operation of the already-built 2G LVK facilities---which primarily requires sustained maintenance---provides a reasonable and conservative baseline. A further aim of this study is to stimulate discussion on long-term strategic investments in gravitational-wave science.


\subsection{Kilonova data}

\begin{table}
\centering
\setlength{\tabcolsep}{12pt}
\renewcommand{\arraystretch}{1.3}
\caption{Number of GW detections and those with kilonova counterparts (GW+KN) under different sky-localization thresholds (\(\Delta\Omega\)) for each detector network, over a 10-year observation period, assuming a simplified Target-of-Opportunity strategy.}
\label{tab:kn_followup}
\begin{tabular}{lccc}
\hline
\textbf{Sky Localization} & \textbf{Detector Network} & \textbf{GW} & \textbf{GW+KN} \\
\hline
\multirow{3}{*}{\vspace{1.5mm}$\Delta \Omega < 10$ deg$^2$} 
  & LVK                & 116     & 23          \\
  & ET+LVK             & 2\,811   & 150   \\
  & CE+ET+LVK          & 38\,621  & 962   \\
\hline
\multirow{3}{*}{\vspace{1.5mm}$\Delta\Omega < 50$ deg$^2$} 
  & LVK                & 177     & 28          \\
  & ET+LVK             & 15\,231  & 548   \\
  & CE+ET+LVK          & 179\,399 & 2\,568 \\
\hline
\multirow{3}{*}{\vspace{1.5mm}$\Delta\Omega < 100$ deg$^2$} 
  & LVK                & 185     & 29          \\
  & ET+LVK             & 26\,060  & 801   \\
  & CE+ET+LVK          & 273\,772 & 3\,183 \\
\hline
\end{tabular}
\end{table}

As discussed above, the \texttt{CosmoDC2\_BCO} synthetic catalog includes electromagnetic counterparts to the gravitational-wave events whenever applicable.  
For neutron star mergers, these counterparts were modeled as kilonovae.  
In the case of BHNS systems, an electromagnetic signal was assigned only when the neutron star was disrupted outside the black hole horizon, while for BNS systems the emission was linked to the nature of the post-merger remnant.  
Luminosities were drawn from ranges consistent with theoretical expectations~\citep{Margalit:2019dpi,Metzger:2019zeh}.  
Apparent magnitudes in the LSST $ugrizy$ filters were then computed by assuming blackbody emission and applying standard $k$-corrections. \footnote{The apparent magnitude in band $b$, compiled in the \texttt{CosmoDC2\_BCO} catalogs, follows the standard formulation of \cite{Hogg:1999ad}. It incorporates the $k$-correction, which accounts for the redshifting of the emitted spectral energy distribution, causing radiation emitted in a given rest-frame band to be mapped onto a different observer-frame band.}

Naturally, the detectable population of kilonovae associated with GW events is subject to several selection effects. The \texttt{CosmoDC2\_BCO} catalogs provide a baseline treatment of electromagnetic detectability by applying the single-visit limiting magnitudes for each LSST band,
\begin{equation}
u = 24.9, \quad
g = 26.0, \quad
r = 25.7, \quad
i = 25.0, \quad
z = 24.3, \quad
y = 23.1 \, .
\label{band_limits}
\end{equation}
These limits correspond to 180s single-visit exposures and follow the recommendations of the Rubin Observatory Target of Opportunity report~\citep{Andreoni:2024pkp}. An event is classified as detectable if its apparent magnitude is brighter than at least one of these thresholds in any band.



In practice, the observable counterpart rate can be further reduced by survey cadence, weather losses, seasonal visibility, and telescope scheduling. Modeling these survey-operational effects would require a dedicated LSST observing-strategy simulation and lies beyond the scope of the present cosmological analysis. To avoid introducing additional model-dependent assumptions, we therefore adopt the conservative choice of using only the photometric detectability criterion already encoded in the catalog.

\begin{figure}
\centering
\includegraphics[width=\textwidth]{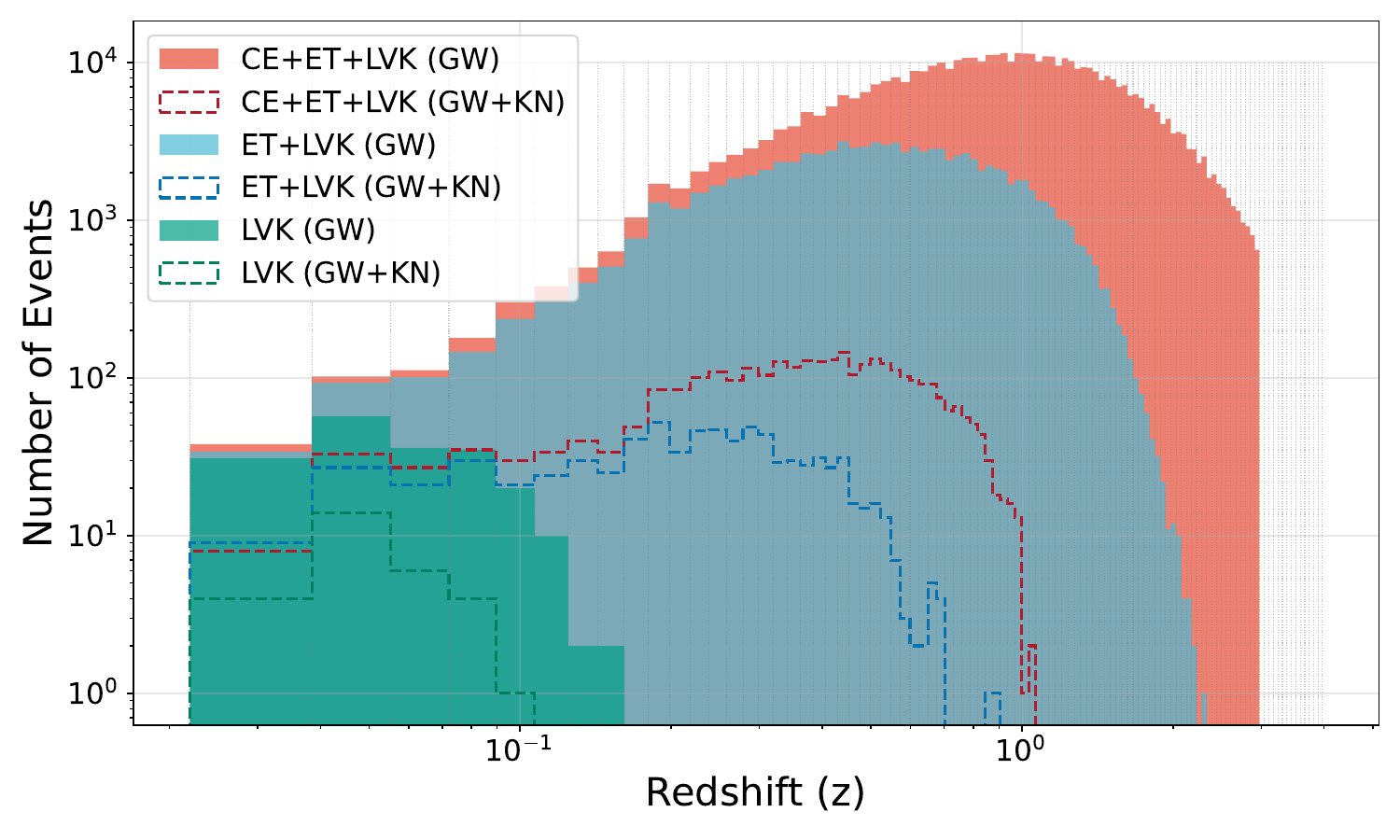}
\caption{Redshift distribution of GW detections (filled histograms) and those with kilonova counterparts (dashed lines) over a 10-year observation period, for different detector networks. }
\label{z_distribution_GW+KN}
\end{figure}

Following a GW trigger, electromagnetic follow-up is not guaranteed due to scheduling, weather, and limited resources. We therefore adopt a simplified selection strategy rather than simulating telescope operations in detail. As a practical proxy for follow-up priority, we use sky-localization precision: events with smaller localization areas require fewer pointings and less observing time, and are thus more likely to be targeted. To bracket plausible strategies, we impose thresholds on the localization area, $\Delta\Omega$, and construct corresponding subsets of GW events likely to receive follow-up. Table~\ref{tab:kn_followup} reports, for each threshold, the total number of GW detections together with the subset that also exhibits kilonova counterparts satisfying the adopted magnitude criteria.

It is also worth noting that, despite a substantial number of BHNS systems being detected in GWs, the released realizations of the \texttt{CosmoDC2\_BCO} catalog yield no electromagnetically detectable counterparts for BHNS when applying the band limits in Eq.~\ref{band_limits}. Consequently, the bright-siren sample used in this work, and summarized in Table~\ref{tab:kn_followup}, consists exclusively of BNS events. For further characterization, Figure~\ref{z_distribution_GW+KN} shows the corresponding redshift distributions, comparing the full GW-detected population with the subset presenting detectable kilonova emission.



In the present analysis, the redshift assigned to each host galaxy corresponds directly to the underlying cosmological redshift provided by the \texttt{CosmoDC2} catalog, and the contribution of peculiar velocities is therefore neglected. As discussed in \cite{Mukherjee:2019qmm,Nimonkar:2023pyt}, the host-galaxy redshift must be accurately corrected for its local motion to avoid biases in $H_0$, particularly for nearby sources. Peculiar velocities correlate with the large-scale density field and with host-galaxy properties such as stellar mass, introducing variations in the inferred Hubble constant of several percent for current detector networks. Proper correction is therefore essential to ensure unbiased and precise cosmological inference from bright sirens.

\subsection{Supernova data}
\label{sec:SN}

\begin{table}
\centering
\setlength{\tabcolsep}{13pt}
\caption{Forecasted number of Type Ia supernovae from the Nancy Grace Roman Space Telescope High-Latitude Time Domain Survey. Counts are separated into deep and wide tiers. The effective statistical uncertainty $\sigma_{\rm eff}$ includes both lensing dispersion and intrinsic scatter, weighted by the number of supernovae per bin.}
\label{tab:roman}
\begin{tabular}{rrrr|rrrr}
\toprule
 $z$ & Deep & Wide & $\sigma_{\rm eff}$ & $z$ & Deep & Wide & $\sigma_{\rm eff}$ \\
\midrule
0.052 &  4 &    0 & 0.0600 & 1.60 & 198 & 214 & 0.0072 \\
0.16 &  14 &   45 & 0.0156 & 1.71 & 184 & 105 & 0.0088 \\
0.26 &  39 &  137 & 0.0091 & 1.81 & 181 &  50 & 0.0101 \\
0.36 &  84 &  214 & 0.0070 & 1.91 & 137 &  17 & 0.0126 \\
0.47 &  83 &  365 & 0.0057 & 2.02 & 119 &   8 & 0.0141 \\
0.57 & 104 &  500 & 0.0050 & 2.12 & 138 &   1 & 0.0137 \\
0.67 & 135 &  683 & 0.0043 & 2.22 & 113 &   1 & 0.0155 \\
0.78 & 155 &  760 & 0.0042 & 2.33 &  96 &   1 & 0.0170 \\
0.88 & 216 &  860 & 0.0039 & 2.43 &  93 &   0 & 0.0177 \\
0.98 & 226 &  982 & 0.0038 & 2.53 &  66 &   0 & 0.0214 \\
1.09 & 214 & 1021 & 0.0038 & 2.64 &  54 &   0 & 0.0240 \\
1.19 & 208 &  930 & 0.0040 & 2.74 &  47 &   0 & 0.0261 \\
1.29 & 206 &  803 & 0.0044 & 2.84 &  35 &   0 & 0.0307 \\
1.40 & 220 &  711 & 0.0046 & 2.95 &  18 &   0 & 0.0435 \\
1.50 & 223 &  453 & 0.0055 &      &     &     &        \\
\bottomrule
\end{tabular}
\end{table}

While GW bright sirens are especially sensitive to the Hubble constant \(H_0\), combining them with complementary probes improves constraints and helps break parameter degeneracies. Type Ia supernovae (SNe Ia) provide strong sensitivity to the matter content and the expansion history. Moreover, bright sirens can anchor the SNe distance scale, enabling an independent assessment of the \(5\sigma\) tension between the \(H_0\) inferred from Planck CMB observations~\citep{Planck:2018vyg} and the Cepheid-calibrated SNe Ia measured by SH0ES~\citep{Riess:2021jrx}.
In this work we include a forecast supernova sample from the Nancy Grace Roman Space Telescope \citep{Rose:2021nzt}, based on the High-Latitude Time Domain Survey. The survey is expected to observe $12\,000$ well-measured Type~Ia supernovae over two years, combining different observing strategies known as wide and deep tiers. The wide tier covers a larger sky area with relatively shallower depth and aims to maximize the total number of detected Type~Ia supernovae at low to intermediate redshifts. In contrast, the deep tier surveys a smaller area with significantly deeper exposures, extending sensitivity to fainter supernovae at higher redshifts. The combination of the two tiers is designed to balance statistical power and redshift reach, providing a well-sampled SN dataset optimized for cosmological analyses.
The adopted Roman-like redshift distribution is reported in Table~\ref{tab:roman}.

The total scatter per supernova combines the intrinsic dispersion and the additional variance from weak lensing:
\begin{equation}
\sigma_{\rm tot}^2(z) = \sigma_{\rm lens}^2(z) + \sigma_{\rm int}^2 ,
\end{equation}
with intrinsic scatter fixed at \(\sigma_{\rm int} = 0.12\) mag.  
The lensing contribution is modeled as~\citep{DES:2024lto}:
\begin{align}
\sigma_{\rm lens}(z) = 0.052 \, \left[ \frac{d_M(z)}{d_M(z=1)} \right]^{3/2} \,,
\end{align}
where the comoving distance to a source at redshift $z$ is:
\begin{align}
d_M(z)=\int_0^z \frac{c\, \d z^\prime}{H(z^\prime)} \,,
\end{align}
and the Hubble function is fixed at the fiducial values.
At low redshift, the intrinsic term dominates, while at higher redshift lensing becomes increasingly important. 

The effective uncertainty on the mean apparent magnitude in each redshift bin is then
\begin{equation}
\sigma_{\rm eff}(z) =
\frac{\sigma_{\rm tot}(z)}{\sqrt{N_{\rm deep}(z) + N_{\rm wide}(z)}} ,
\end{equation}
where $N_{\rm deep}$ and $N_{\rm wide}$ are the numbers of SNe in the Roman deep and wide tiers. Each redshift bin is treated as a single effective measurement, reducing the dataset to 29 points and thereby improving numerical efficiency. All supernovae are assumed to be independent, yielding a diagonal covariance matrix.

Apparent magnitudes, \( m_B(z) \), are computed from the distance modulus $\mu(z)$ relative to the \texttt{CosmoDC2} fiducial cosmology, assuming an absolute magnitude \( M_B = -19.3 \):
\begin{align} \label{mB}
\mu(z) &= 5 \log_{10}\! \frac{d_L(z)}{10\,\mathrm{pc}} , \\
m_B(z) &= \mu(z) + M_B \,,
\end{align}
where \( d_L(z) = (1+z)\,d_M(z) \) is the luminosity distance.

\section{Methodology}
\label{Methodology}

\subsection{Cosmological models}
\label{Methodology:Cosmo_Models}

Our goal is to explore how bright sirens can constrain key features of the cosmic expansion, from the Hubble constant to the possible dynamics of dark energy. For this reason, we test a representative set of cosmological scenarios:

\begin{itemize}
\item Flat \(\Lambda\)CDM: serves as the baseline model for most cosmological analyses, providing a reference to quantify the constraining power of bright sirens and supernovae. Its Hubble function is $H^2(z)/H^2_0=\Om  \, (1+z)^{3} + 1- \Om $.

\item Flat \(w\)CDM: allows testing for deviations from a cosmological constant. Its Hubble function is $H^2(z)/H^2_0=\Om  \, (1+z)^{3} + 1- \Om \, (1+z)^{3 (1+w_0)}$.

\item Flat \(w_0w_a\)CDM: captures possible time evolution in the dark energy equation of state. Its Hubble function is $H^2(z)/H^2_0=\Om  \, (1+z)^{3} + 1- \Om \, (1+z)^{3 q(z)}$, where, for the CPL parameterization $w(z)=w_{0}+ w_a {z \over 1+z}$, it is $q(z)= 1+w_{0}+w_{a}- {w_{a} \, z \over (1+z)\ln (1+z)}$~\citep{Chevallier:2000qy,Linder:2002et}.

\item Cosmography: enables model-independent tests of the expansion history at low redshift, useful for directly comparing with local measurements of $H_0$. We adopt, as in \cite{Riess:2021jrx}, a second-order expansion of the luminosity distance \cite{Camarena:2023rsd}:
\begin{align} \label{dLcg}
d_L(z) &=  \frac{c z}{H_0} \left [ 1 + \frac{1-q_0}{2}\, z  -\frac{1- q_0-3 q_0^2+j_0 -\Omega_{k0} }{6} \, z^2  + O(z^3) \right] \,,
\end{align}
where the Hubble constant $H_0$, the deceleration parameter $q_0$ and the jerk parameter $j_0$ are defined, respectively, according to:
\begin{align}
H_0=\left. \frac{\dot a(t)}{a(t)} \right|_{t_0} \,,
\qquad
q_0=\left. \frac{- \ddot a(t)}{H^2(t) a(t)} \right|_{t_0}  \,,
\qquad
j_0=\left.\frac{\dddot a(t)}{H^3(t) a(t)}\right|_{t_0} \,.
\end{align}
For the fiducial $\Lambda$CDM model, $q_0$ and $j_0$ are given by:
\begin{align}
q_0 =\frac{3}{2} \Omega_{\rm m 0} -1
\stackrel{\text{fid}}{=} -0.6028 \,, \qquad
j_0  =  1 \,. 
\end{align}

\end{itemize}

\subsection{Time-evolution of constraints}

Gravitational wave observatories are designed to operate over extended periods. Consequently, cosmological inference will naturally evolve as more events are detected. To account for this, we evaluate the constraints as a function of observation time.

The number of mergers is assumed to scale linearly with the observing time. For an effective duration \(t_{\rm obs}\), we therefore select a fraction \(t_{\rm obs}/10\,{\rm yr}\) of the full GW catalog, which was originally generated for a ten-year observing period. This scaling neglects possible deviations arising from varying detector duty cycles, and should thus be regarded as a simplifying assumption adopted for forecasting purposes. The same procedure is applied to the SNe sample, scaled to its nominal two-year duration. When combining GW and SNe data, a key consideration is whether the observations are contemporaneous. The Roman Space Telescope and the LVK O5 run are expected to overlap in time, which justifies using only the corresponding fractions of each dataset. In contrast, 3G detectors such as CE and ET are expected to operate well after Roman. Therefore, in analyses involving 3G GW events and SNe data, we always use the full SNe sample, while the \(t_{\rm obs}\)-dependent scaling is applied only to the GW data.

In this analysis we focus primarily on the evolution of the Hubble constant constraint, as \(H_0\) is a key parameter across all cosmological models considered. For models that allow dynamical dark energy, we also track the Figure of Merit (FoM), which quantifies the joint constraining power on the equation-of-state parameters \(w_0\) and \(w_a\). Following the standard Dark Energy Task Force (DETF) convention~\citep{Albrecht:2006um}, the FoM is defined as
\begin{equation} \label{fom}
    \mathrm{FoM} = \frac{1}{\sqrt{\det(C_{w_0, w_a})}}\,,
\end{equation}
where \(C_{w_0, w_a}\) is the covariance matrix of \(w_0\) and \(w_a\). This definition makes explicit that the FoM is inversely proportional to the area of the uncertainty ellipse in the \(w_0\)–\(w_a\) plane, providing a direct measure of the precision with which dynamical dark energy models can be constrained.

\subsection{Likelihood Construction and Priors}
\label{sec:like}

\begin{table}
\centering
\setlength{\tabcolsep}{16pt}
\caption{Priors for the cosmological parameters.}
\label{priors}
\begin{tabular}{ll}
\toprule
\textbf{Parameter} & \textbf{Prior range} \\
\midrule
\(H_0\) & Uniform in \([40, 120]\, \mathrm{km\,s^{-1}\,Mpc^{-1}}\) \\
\(\Omega_m\) & Uniform in \([0.0, 0.9]\) \\
\(w_0\) & Uniform in \([-10, 0]\) \\
\(w_a\) & Uniform in \([-10, 10]\) \\
\(M_B\) & Uniform in \([-20.5, 18.5]\) mag \\
\bottomrule
\end{tabular}
\end{table}

Bayesian inference is performed using the \texttt{emcee} sampler \citep{ForemanMackey:2012ig}, and the Table~\ref{priors} summarizes the uniform priors adopted for the cosmological parameters. In particular, the prior on \(M_B\) is chosen to be sufficiently broad so as not to restrict the MCMC exploration, while comfortably encompassing the fiducial value \(M_B \simeq -19.3\). This reference value is consistent with recent SH0ES calibrations of Type~Ia supernovae \cite{Riess:2021jrx}, but no informative prior  is imposed. 
Posterior samples are used to derive marginalized constraints and are visualized using the \texttt{GetDist} package \citep{Lewis:2019xzd}.

To infer cosmological parameters from the simulated observations, we construct a total likelihood function that combines both gravitational-wave and supernova datasets. Since the two datasets are statistically independent, the total likelihood is given by the product of the individual likelihoods, or equivalently, by the sum of their log-likelihoods:
\begin{equation}
\ln \mathcal{L}_{\mathrm{total}} = \ln \mathcal{L}_{\mathrm{GW}} + \ln \mathcal{L}_{\mathrm{SNe}}
\end{equation}
A Gaussian likelihood is assumed in both cases. For analyses that use only GW data, the second term is  omitted. Each contribution is described in the following sections.

\subsubsection{Bright sirens likelihood}

We probe the expansion history through the forecasted constraints on the luminosity distance from GW events.  Considering the Fisher matrix $\Gamma^{\rm (net)}$ for the $i$-th event detected by a given  network (see Eq.~\ref{fisher_net}), the marginalized uncertainty on the luminosity distance is obtained as
\begin{equation}
\sigma_{d_L,i} = \sqrt{\mathcal{C}^{\rm (net)}_{d_L d_L}},
\end{equation}
where the network covariance matrix is given by $\mathcal{C}^{\rm (net)} = \left(\Gamma^{\rm (net)}\right)^{-1}$.  
Denoting by $d_{L,i}^{\mathrm{GW}}$ the fiducial (true) value of luminosity distance of the $i$-th event, the GW cosmological likelihood can then be written as
\begin{equation}
\ln \mathcal{L}_{\mathrm{GW}} = -\frac{1}{2} \sum_i 
\left[
\frac{ d_L(z_i;\boldsymbol{\theta}_{\mathrm{cosmo}}) - d_{L,i}^{\mathrm{GW}} }
     { \sigma_{d_L,i} }
\right]^2 ,
\label{gw_likelihood}
\end{equation}
where $z_i$ is the redshift of the $i$-th event and $\boldsymbol{\theta}_{\mathrm{cosmo}}$ denotes the set of cosmological parameters to be constrained.

To ensure robustness, we include only those gravitational-wave events whose covariance matrices are flagged as numerically stable, based on internal diagnostics from the simulation pipeline \citep{Menote:2025zmn}. Events with non-invertible or ill-conditioned Fisher matrices are excluded from the analysis.

However, for the LVK configuration this approximation becomes less reliable.  
Events detected by the LVK network are affected by larger measurement uncertainties, and in this regime the Gaussian assumption underlying the Fisher approach may break down. In particular, the physical constraint \( d_L > 0 \) can truncate the posterior distribution, leading to asymmetric shapes and potential biases.  
To address this, we adopt a different strategy for the LVK case. Owing to the relatively small number of detected events (23, see Table~\ref{tab:kn_followup}), we perform a dedicated MCMC analysis for each event. This approach allows us to explore the full posterior distribution without relying on the Fisher approximation. The marginalized distance uncertainties, \( \sigma_{d_L,i} \), are then obtained directly as the standard deviation of the posterior samples.

\subsubsection{Supernovae likelihood}

The supernova likelihood is written as:
\begin{align}
\ln \mathcal{L}_{\mathrm{SNe}} = -\frac{1}{2} \sum_i \left[ \frac{m_B(z_i;\theta_{\mathrm{cosmo}}) -  m_{B,i}}{\sigma_{{\rm eff},i}} \right]^2 ,
\end{align}
where the apparent magnitudes $m_{B,i}$, redshifts $z_i$ and uncertainties \(\sigma_{{\rm eff},i}\) were discussed in Section~\ref{sec:SN}.
When not interested in the absolute magnitude \( M_B \), it is computationally efficient to marginalize analytically over it, yielding the following effective likelihood~\citep{Camarena:2021jlr}:
\begin{align}
\ln \mathcal{L}_{\mathrm{SNe}}^{\rm marg} = -\frac{1}{2} \left ( S_2 - \frac{S_1^2}{S_0} \right) \,,
\end{align}
where
\begin{align}
\delta_{\mathrm{SNe},i} \equiv m_{B,i} - \mu(z_i; \theta_{\text{cosmo}}) \,, \quad
S_0 = \sum_i \frac{1}{\sigma_{{\rm eff},i}^2} \,, \quad
S_1 = \sum_i \frac{\delta_{\mathrm{SNe},i}}{\sigma_{{\rm eff},i}^2} \,, \quad
S_2 = \sum_i  \frac{\delta_{\mathrm{SNe},i}^2}{\sigma_{{\rm eff},i}^2} \,.
\end{align}

\section{Results}
\label{sec:results}

\begin{table}
\centering
\setlength{\tabcolsep}{5pt}
\renewcommand{\arraystretch}{1.2}
\caption{Marginalized parameter constraints (68\% credible intervals) for different cosmological models and detector networks after ten years of observation. Results are shown for bright sirens only (BS only) and for the joint analysis with supernovae from Roman (BS+SNe).}
\label{tab:constraints}
\resizebox{\textwidth}{!}{%
\begin{tabular}{ccccccccc}
\toprule
Model & Network & Data & $H_0  \left( \frac{\rm km/s}{\rm Mpc} \right)$ & $\frac{\sigma_{H_0}}{H_0}$[\%] & $\Omega_m$ & $w_0$ & $w_a$ & $M_B$ (mag) \\
\midrule
\multirow{6}{*}{\parbox{2.8cm}{\centering Cosmography\\ \(0.023\!<\!\!z\!\!<\!0.15\) \\ ($q_0 = -0.603$ \\and $j_0 = 1$)}}
 & \multirow{2}{*}{LVK} 
   & BS only   & $71.0 ^{+4.2}_{-3.8}$ & 5.6 & --- & --- & --- & --- \\
 &   & BS+SNe  & $71.0 ^{+4.4}_{-3.9}$ & 5.8 & --- & --- & --- & $-19.30 ^{+0.13}_{-0.12}$ \\
 \cmidrule(lr){2-9}
 & \multirow{2}{*}{ET+LVK} 
   & BS only   & $71.00 \pm 0.23$ & 0.32 & --- & --- & --- & --- \\
 &   & BS+SNe  & $71.00 \pm 0.22$ & 0.31 & --- & --- & --- & $-19.300 ^{+0.017}_{-0.016}$ \\
 \cmidrule(lr){2-9}
 & \multirow{2}{*}{CE+ET+LVK} 
   & BS only   & $70.99 \pm 0.15$ & 0.21 & --- & --- & --- & --- \\
 &   & BS+SNe  & $71.00 \pm 0.15$ & 0.21 & --- & --- & --- & $-19.299 ^{+0.015}_{-0.016}$ \\
\midrule
\multirow{6}{*}{$\Lambda$CDM}
 & \multirow{2}{*}{LVK}
   & BS only   & $72.0 \pm 3.9$ & 5.4 & unconstr. & --- & --- & --- \\
 &   & BS+SNe & $71.0 \pm 3.8$ & 5.4 & $0.2648 \pm 0.0031$ & --- & --- & marg. \\
 \cmidrule(lr){2-9}
 & \multirow{2}{*}{ET+LVK}
   & BS only   & $71.00 \pm 0.34$ & 0.48 & $0.265 \pm 0.062$ & --- & --- & --- \\
 &   & BS+SNe & $71.00 \pm 0.20$ & 0.28 & $0.2648 \pm 0.0030$ & --- & --- & marg. \\
 \cmidrule(lr){2-9}
 & \multirow{2}{*}{CE+ET+LVK}
   & BS only   & $71.00 \pm 0.17$ & 0.24 & $0.265 \pm 0.015$ & --- & --- & --- \\
 &   & BS+SNe & $71.00 \pm 0.12$ & 0.17 & $0.2648 \pm 0.0029$ & --- & --- & marg. \\
\midrule
\multirow{6}{*}{$w$CDM} 
 & \multirow{2}{*}{LVK} 
   & BS only   & $71.0 ^{+5.1}_{-4.4}$ & 6.7 & unconstr. & unconstr. & --- & --- \\
 &   & BS+SNe  & $71.0 ^{+4.1}_{-3.6}$ & 5.4 & $0.2648 ^{+0.0049}_{-0.0053}$ & $-0.997 ^{+0.036}_{-0.037}$ & --- & marg. \\
 \cmidrule(lr){2-9}
 & \multirow{2}{*}{ET+LVK} 
   & BS only   & $71.00 ^{+0.59}_{-0.46}$ & 0.74 & $0.26 ^{+0.16}_{-0.24}$ & $-1.02 ^{+0.54}_{-1.04}$ & --- & --- \\
 &   & BS+SNe  & $70.99 \pm 0.23$ & 0.32 & $0.2648 \pm ^{+0.0050}_{-0.0052}$ & $-1.00 ^{+0.04}_{-0.04}$ & --- & marg. \\
 \cmidrule(lr){2-9}
 & \multirow{2}{*}{CE+ET+LVK} 
   & BS only   & $70.99 ^{+0.27}_{-0.25}$ & 0.37 & $0.275 ^{+0.083}_{-0.110}$ & $-1.00 ^{+0.19}_{-0.21}$ & --- & --- \\
 &   & BS+SNe  & $70.99 \pm 0.16$ & 0.23 & $0.2648 ^{+0.0046}_{-0.0047}$ & $-1.003 ^{+0.028}_{-0.030}$ & --- & marg. \\
\midrule
\multirow{6}{*}{$w_0w_a$CDM} 
 & \multirow{2}{*}{LVK} 
   & BS only   & $70.8 ^{+7.5}_{-5.3}$ & 9.0 & unconstr. & unconstr. & unconstr. & --- \\
 &   & BS+SNe  & $71.0 ^{+4.2}_{-3.8}$ & 5.6 & $0.266 ^{+0.027}_{-0.056}$ & $-1.00 ^{+0.94}_{-0.07}$ & $-0.03 ^{+0.72}_{-0.69}$ & marg. \\
 \cmidrule(lr){2-9}
 & \multirow{2}{*}{ET+LVK} 
   & BS only   & $70.99 ^{+0.94}_{-0.78}$ & 1.2 & $0.396 ^{+0.073}_{-0.160}$ & $-1.2 ^{+0.8}_{-1.4}$ & $-1 ^{+13}_{-17}$ & --- \\
 &   & BS+SNe  & $71.01 ^{+0.31}_{-0.34}$ & 0.46 & $0.265^{+0.025}_{-0.071}$ & $-1.00x ^{+0.081}_{-0.063}$ & $0.00 ^{+0.74}_{-0.60}$ & marg. \\
 \cmidrule(lr){2-9}
 & \multirow{2}{*}{CE+ET+LVK} 
   & BS only   & $70.99 ^{+0.36}_{-0.40}$ & 0.54 & $0.27 ^{+0.08}_{-0.16}$ & $-0.99 ^{+0.19}_{-0.21}$ & $-0.0 ^{+2.7}_{-5.2}$ & --- \\
 &   & BS+SNe  & $71.00 \pm 0.23$ & 0.32 & $0.265 ^{+0.016}_{-0.024}$ & $-1.001 ^{+0.040}_{-0.037}$ & $0.01 ^{+0.39}_{-0.37}$ & marg. \\
\bottomrule
\end{tabular}
}
\end{table}

\subsection{Local measurement of $H_0$}

We perform a SH0ES-like analysis~\citep{Riess:2021jrx}, restricting to the redshift range \(0.023 < z < 0.15\). The lower cutoff suppresses most of the cosmic variance in $H_0$ due to local large-scale structure~\citep{Camarena:2018nbr}, while the upper cutoff ensures that the measurement remains local and that the cosmographic expansion of Section~\ref{Methodology:Cosmo_Models} is valid. 
In this regime, we fix the deceleration and jerk parameters, $q_0$ and $j_0$, to their fiducial values, following~\cite{Riess:2021jrx}, and fit only for $H_0$. When supernova data are included, we also fit for the absolute magnitude $M_B$, thereby calibrating the supernova luminosity scale using bright siren observations. This offers an important consistency test of the traditional Cepheid-based calibration of Type Ia supernovae~\citep{Riess:2021jrx}, which underlies the well-known \(5\sigma\) tension with the $H_0$ and $M_B$ \citep{Camarena:2023rsd} values inferred from Planck CMB data~\citep{Planck:2018vyg}.

Table~\ref{tab:constraints} summarizes the constraints obtained for the three detector networks considered in this work.  
As expected, the precision on $H_0$ improves significantly when moving from 2G to 3G networks, shrinking from about 6\% to  0.3\%. This gain arises from the enhanced sensitivity and extended horizon of 3G detectors, which provide both a larger number of bright-siren detections and better distance precision per event.  
For the LVK case, we conclude that the O5 observing run has the potential to shed light on the $H_0$ tension \textit{if} the measured value does not fall within the Planck and SH0ES determinations, which currently achieve precisions of 0.8\% and 1.4\%, respectively.  
The addition of CE to the ET+LVK network brings only a modest further improvement. Nonetheless, additional gains may be achieved through refined physical modeling of the very high-SNR events that the CE+ET+LVK configuration will detect in the nearby Universe.

The inclusion of supernovae, as expected, does not impact the measurement of $H_0$. Instead, the synergy with bright sirens enables the absolute calibration of the supernova absolute magnitude. While the LVK network cannot compete with the Cepheid-based calibration by SH0ES, which achieves \(\sigma_{M_B} = 0.027\)~mag~\citep{Riess:2021jrx}, the 3G networks demonstrate the potential to constrain $M_B$ with uncertainties nearly a factor of two tighter than the current best measurements. This establishes GW bright sirens as a powerful cornerstone for anchoring the cosmic distance ladder in the coming decades.

\subsection{Cosmological parameter estimation}
\label{subsec:bs_constraints}

\begin{figure}[p]
\centering
\begin{minipage}{0.49\textwidth}
  \includegraphics[width=0.62\linewidth]{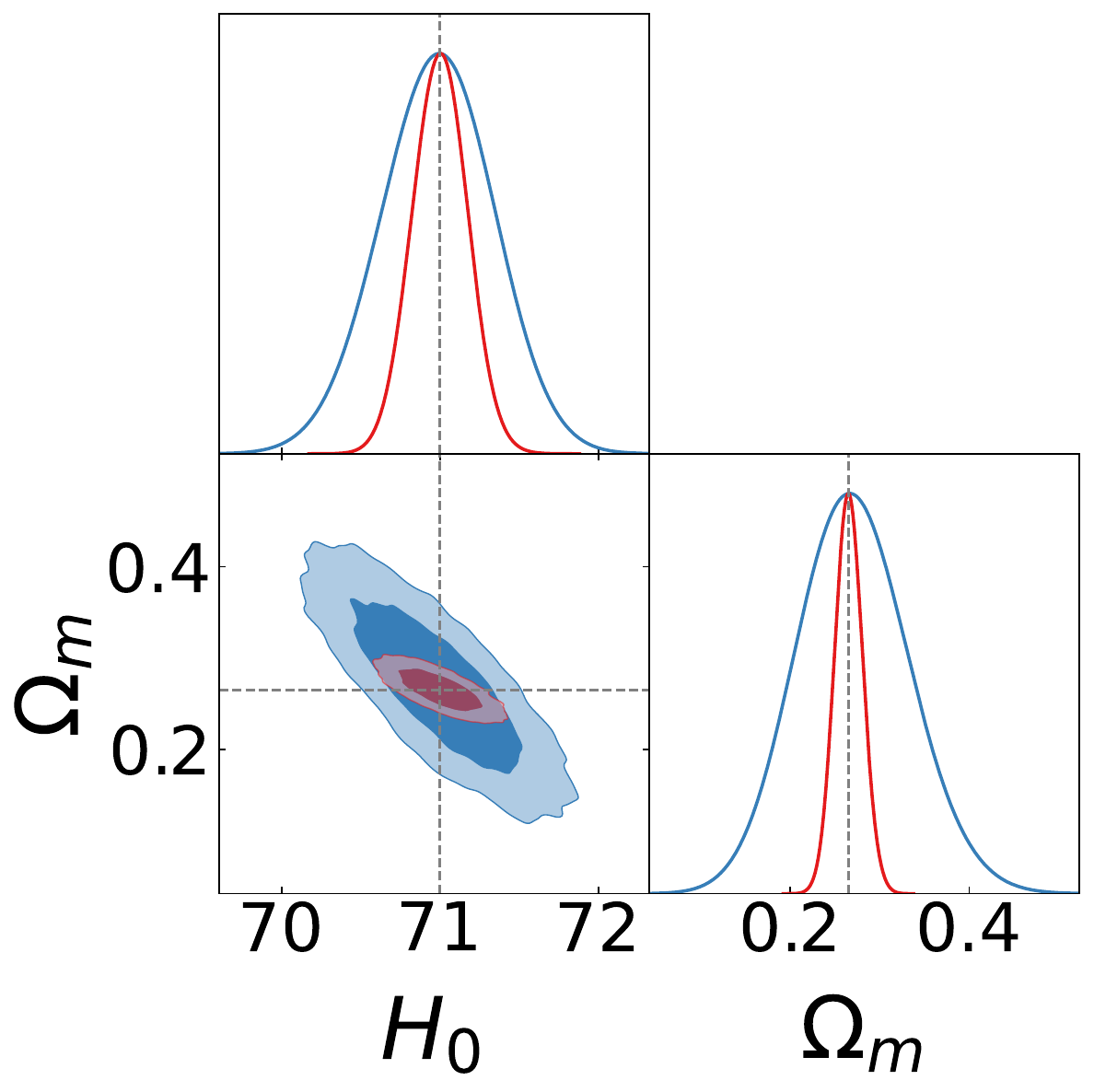}
  \vspace{1em}
  \includegraphics[width=0.82\linewidth]{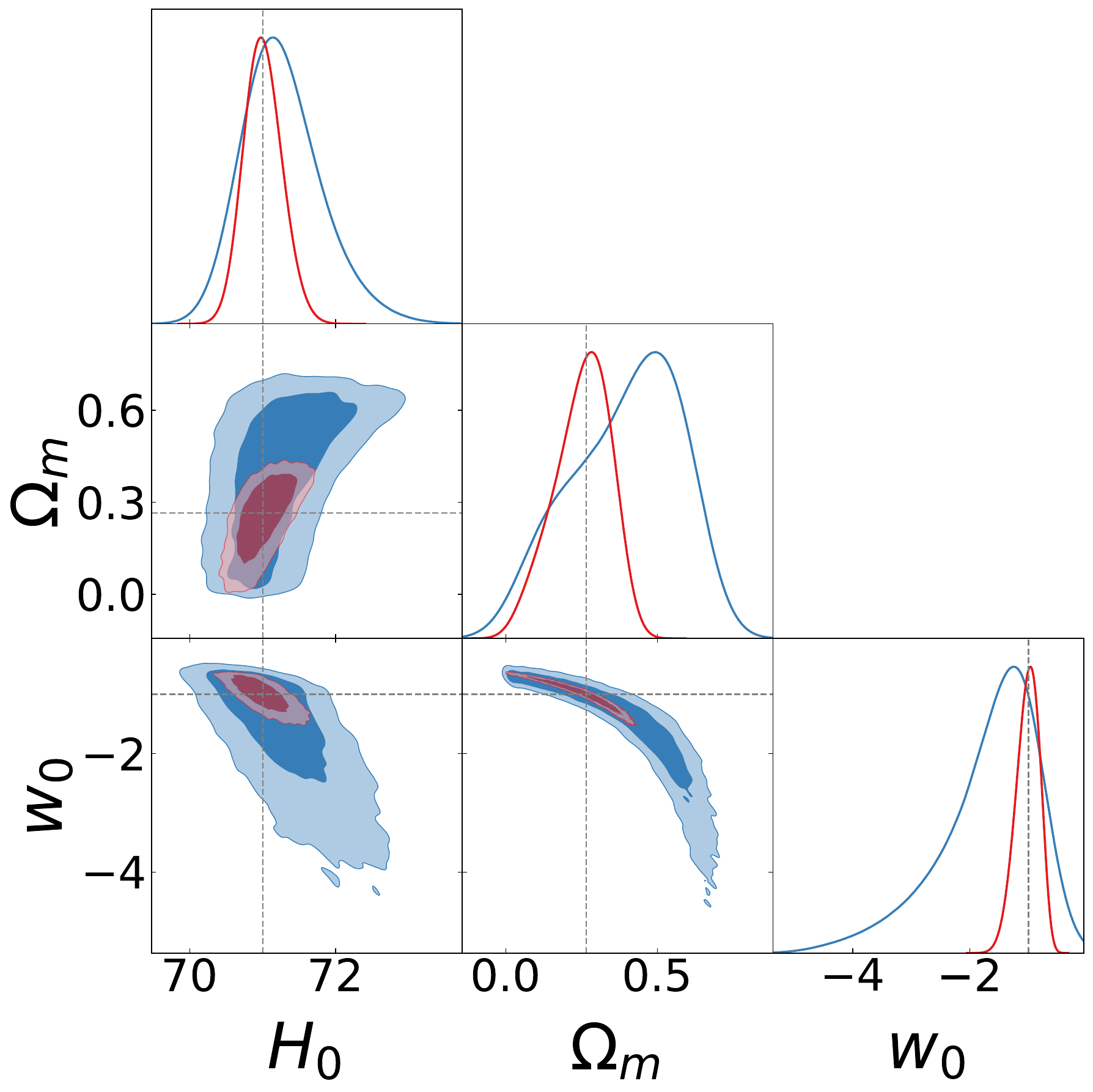}
  \vspace{1em}
  \includegraphics[width=\linewidth]{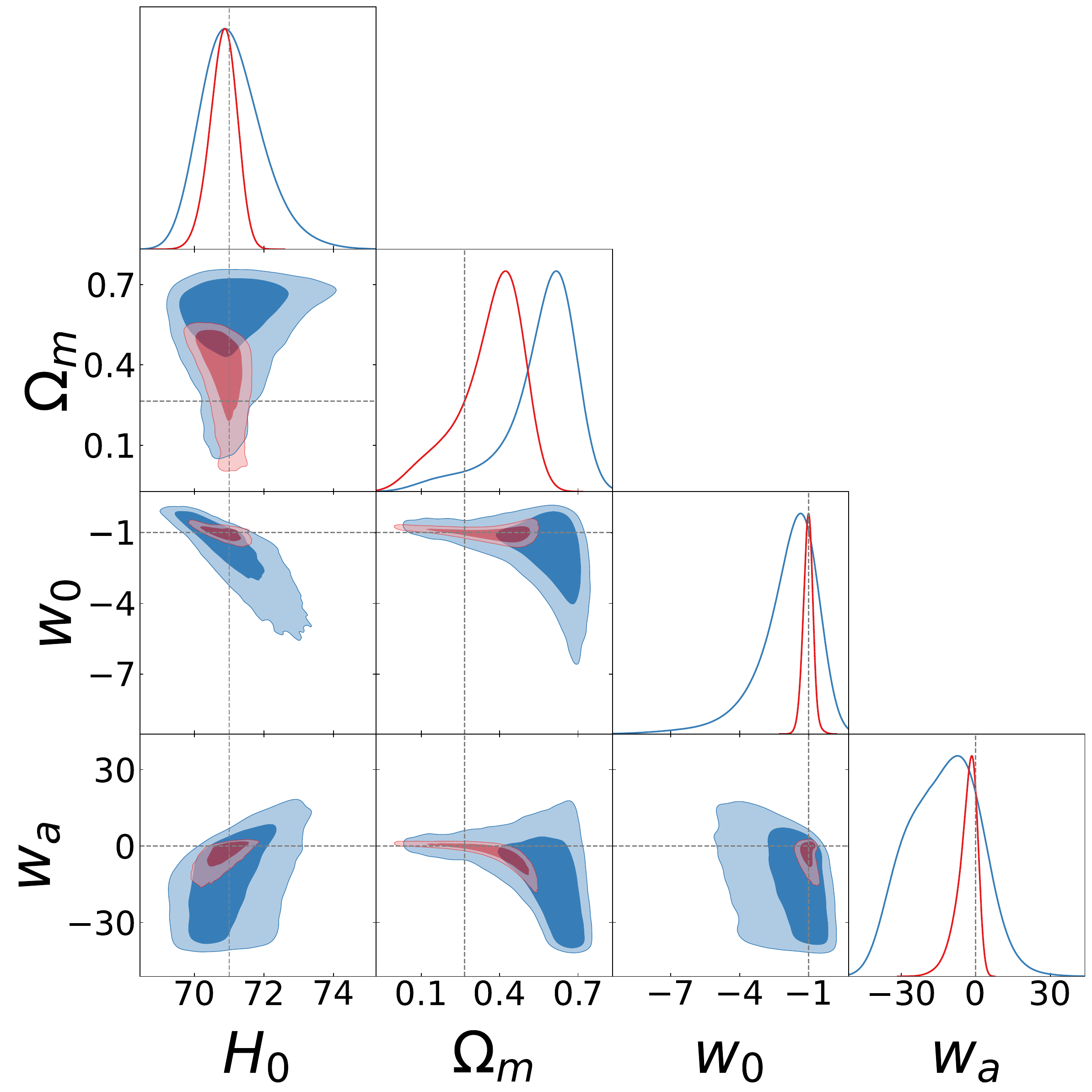}
\end{minipage}%
\hfill
\begin{minipage}{0.49\textwidth}
  \includegraphics[width=0.62\linewidth]{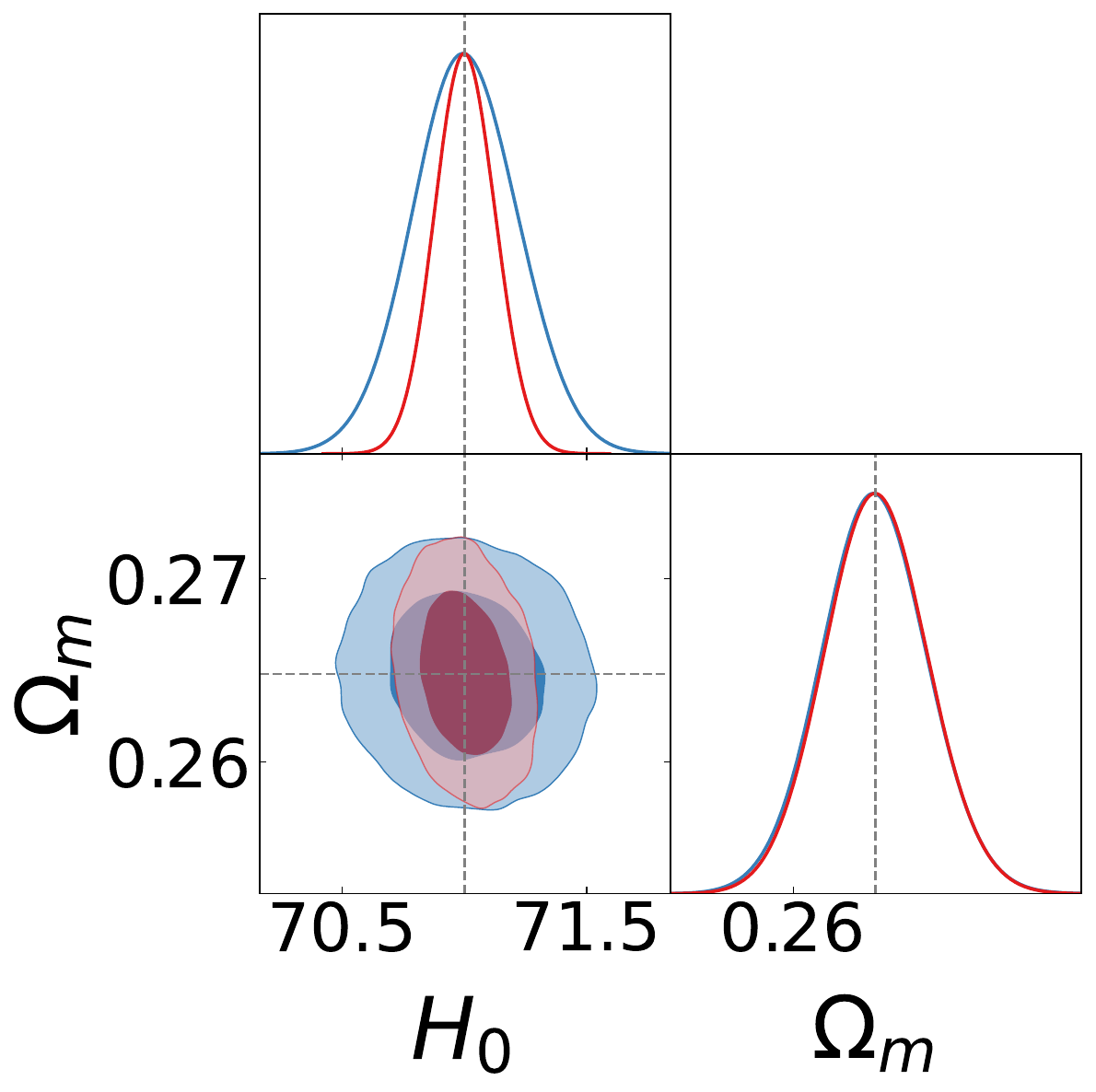}
  \vspace{1em}
  \includegraphics[width=0.82\linewidth]{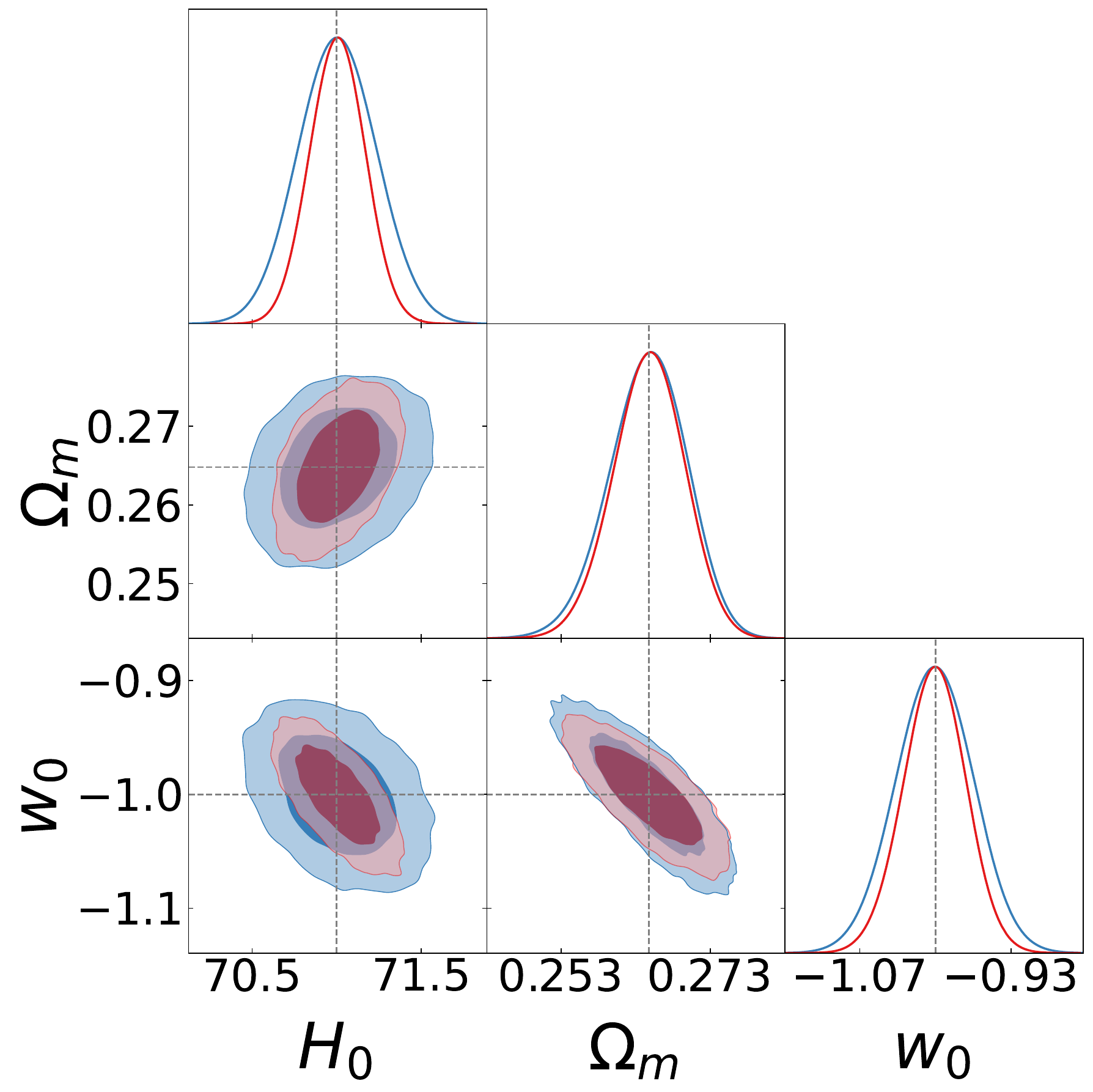}
  \vspace{1em}
  \includegraphics[width=\linewidth]{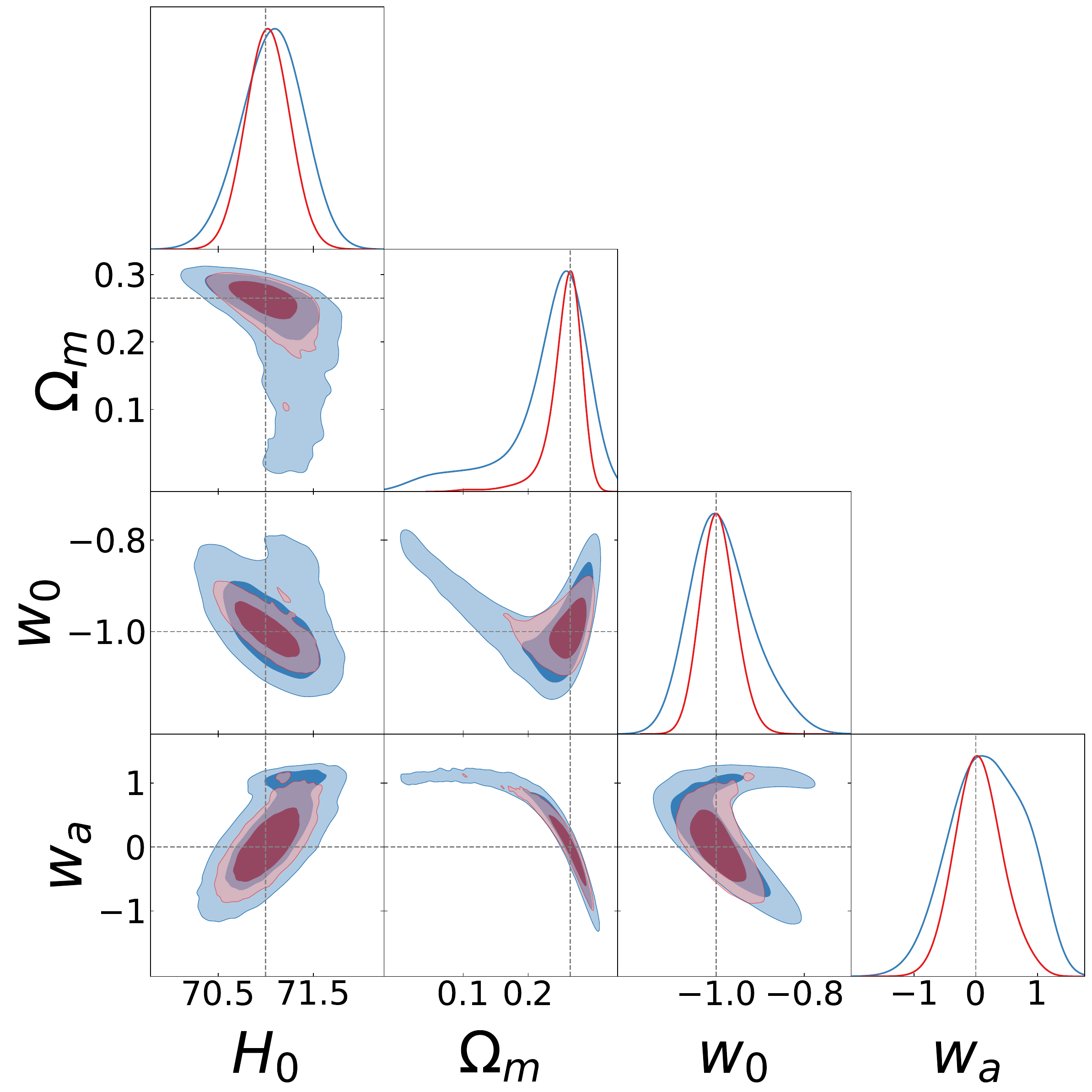}
\end{minipage}
\caption{Comparison of cosmological constraints from BS only (left) and BS + SNe (right) for 3G detector networks ET+LVK (blue) and CE+ET+LVK (red) 
The analysis includes GW events with sky localization better than \(100\,\mathrm{deg}^2\) over a total observation time of 10 years. The top row corresponds to $\Lambda$CDM, the second to $w$CDM, and the bottom row to $w_0w_a$CDM. The fiducial parameter values are indicated by the dashed lines.}
\label{fig:network_posteriors}
\end{figure}

Table~\ref{tab:constraints} and Figure~\ref{fig:network_posteriors} present the forecasted constraints for the $\Lambda$CDM, $w$CDM, and $w_0w_a$CDM models. Third-generation networks yield subpercent precision on $H_0$, establishing bright sirens as a definitive probe of the Hubble constant. In the more flexible $w_0w_a$CDM model, BS-only constraints on $H_0$ are already comparable to those from Planck (0.8\%); adding Roman supernovae reduces the uncertainty to  below 0.5\%.  
Although the unanchored supernova dataset does not directly constrain $H_0$, it indirectly strengthens its inference by propagating additional constraining power through parameter correlations.

In contrast, the constraints on $\Omega_m$ do not show the same improvement in precision. In the $\Lambda$CDM case, bright sirens alone add little to the information already provided by supernovae; even with the inclusion of 3G detectors, the reduction in the forecasted uncertainty remains marginal. This weaker sensitivity arises from the redshift distribution of the bright sirens (Figure~\ref{z_distribution_GW+KN}), which is concentrated at low redshift, where the expansion history is already dominated by dark energy and thus less sensitive to the matter density parameter.

Regarding dark energy, Table~\ref{tab:constraints} shows that bright sirens probe its time evolution and significantly tighten the constraints on $w_0$ and $w_a$, best summarized by the Figure of Merit defined in Eq.~\eqref{fom}. Table~\ref{tab:FoM_comparison} reports the FoM values obtained for 3G detector networks combined with Roman supernovae and compares them with the latest DESI analysis~\cite{DESI:2025zgx}. The DESI FoM values were computed from the publicly available posterior chains~\cite{desi_collaboration_2025_16644577}.
Notably, the CE+ET+LVK plus Roman configuration attains a FoM surpassing that of DESI
DR2 BAO combined with DES Year 5 supernovae.

Finally, Figure~\ref{fig:network_posteriors} shows that bright-siren data alone have a limited ability to reconstruct the matter content of the Universe. In the $w$CDM model, a meaningful constraint on $\Omega_m$ is obtained only when two third-generation detectors are combined: the ET+LVK network returns a value of $\Omega_m$ that is displaced by roughly one standard deviation from the fiducial value, likely due to volume effects. In the more flexible $w_0w_a$CDM model, not even the CE+ET+LVK configuration is able to recover the fiducial parameters, clearly highlighting the need for synergy with supernova data.

\begin{table}
\centering
\setlength{\tabcolsep}{25pt}
\renewcommand{\arraystretch}{1.3}
\caption{Comparison of the Figure of Merit (FoM) in the $w_0$-$w_a$ plane for different datasets.}
\label{tab:FoM_comparison}
\begin{tabular}{lc}
\toprule
Dataset  & FoM \\
\midrule
DESI DR2 + DESY5 \cite{DESI:2025zgx}       & 55.9 \\
DESI DR2 + CMB + DESY5 \cite{DESI:2025zgx} & 185.9 \\
\midrule
ET + LVK (10 yr) + Roman (2 yr)   & 24.9 \\
CE + ET + LVK (10 yr) + Roman (2 yr) & 76.1 \\
\bottomrule
\end{tabular}
\end{table}

\subsection{Time evolution of constraints}
\label{subsec:time_evolution_constraints}

\begin{figure}[p]
\centering
\includegraphics[width=\textwidth]{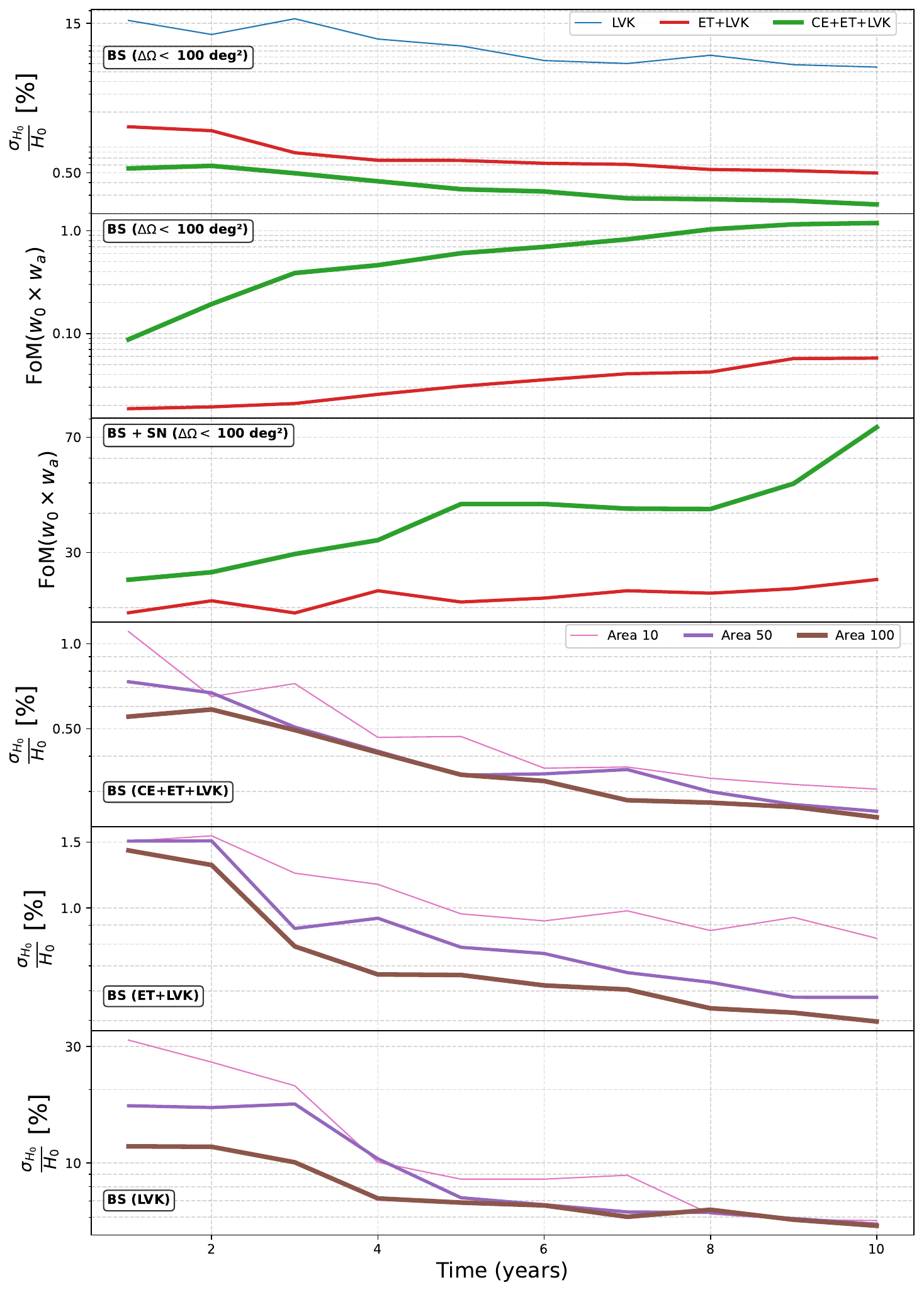}
\caption{Time evolution of $\sigma_{H_0}$ for the $\Lambda$CDM model using BS only (top panel), and of the Figure of Merit on $w_0$ and $w_a$ for BS only (second panel) and BS + SNe (third panel), restricted to events with sky localization better than \(100\,\mathrm{deg}^2\). The bottom three panels show the time evolution of $\sigma_{H_0}$ under different sky-localization thresholds.}
\label{fig:time_evolution}
\end{figure}

Figure~\ref{fig:time_evolution} (top panel) shows the evolution of the Hubble constant uncertainty as a function of observation time for the $\Lambda$CDM model, considering bright sirens with sky localization better than \(100\,\mathrm{deg}^2\). 
Third-generation networks substantially accelerate the path to high-precision cosmology. The CE+ET+LVK configuration rapidly accumulates well-localized events, driving $\sigma_{H_0}$  to 0.6\%  within the first two years and reaching 0.2\%  after a decade. This precision surpasses current results from traditional cosmological probes and is achieved without reliance on external distance ladders. The ET+LVK configuration attains a similar level, reaching $\sigma_{H_0} \approx 0.6\%$ in roughly six years.  
By contrast, the current-generation LVK network, which represents the realistic near-term scenario, converges much more slowly: six years are necessary to go below 7\%.

The time evolution of the Figure of Merit is displayed in the second and third panels of Figure~\ref{fig:time_evolution}. The addition of supernova data substantially enhances the constraining power when combined with bright sirens. In this case, the FoM grows almost linearly with observation time, reflecting the steady accumulation of well-measured events. Reaching the level achieved by DESI DR2 BAO together with DES Year 5 supernovae requires nearly a full decade of observations.

Finally, the last three panels of Figure~\ref{fig:time_evolution} show the evolution of the Hubble-constant uncertainty as a function of observing time in the $\Lambda$CDM model, for three detector networks and for the sky--localization cuts listed in Table~\ref{tab:kn_followup} (\(\Delta\Omega<10\), \(50\), \(100~\mathrm{deg}^2\)). The aim is to bracket realistic EM follow-up scenarios.
For LVK, precision improves slowly and the curves display pronounced fluctuations caused by small-number statistics. Only after \(\sim\!8\) years do the different localization cuts converge to comparable precision (\(\sim\!6\%\)). This behavior reflects  the fact that poorly localized events typically carry larger distance uncertainties and therefore add little statistical weight.
For 3G networks, the penalty from stricter localization is modest for \(\Delta\Omega<50~\mathrm{deg}^2\) but becomes more evident for \(\Delta\Omega<10~\mathrm{deg}^2\). In both cases, however, the uncertainty falls below the 1\% and 0.5\% levels after about five years for ET+LVK and CE+ET+LVK, respectively. These results suggest that EM follow-up can be optimized by prioritizing events localized within \(\sim\!50~\mathrm{deg}^2\), or more aggressively \(\sim\!10~\mathrm{deg}^2\), which preserves significant cosmological information while considerably accelerating ToO campaigns.

\section{Conclusion}
\label{sec:conclusions}

In this work, we investigated the cosmological constraining power of bright sirens and their synergy with the future Roman Type Ia supernova dataset. Using the \mytt{CosmoDC2_BCO} catalog of events detectable by current and next-generation GW detector networks, combined with kilonova discoveries enabled by LSST Target of Opportunity observations, we assessed how the number of events, sky-localization accuracy, and joint observations with SNe~Ia affect the precision of cosmological parameters.

For the Hubble constant, using BNS/BHNS events localized within \(\Delta\Omega \lesssim 100~\mathrm{deg}^2\), third-generation networks achieve sub-percent precision after only three years of observations, reaching 0.2\% with the CE+ET+LVK configuration after a decade. By contrast, the LVK network requires nearly ten years to reduce the uncertainty below 6\%. This implies that the LVK O5 observing run can shed light on the $H_0$ tension \textit{only if} the inferred value lies outside the range spanned by the Planck and SH0ES determinations, which currently achieve much higher precisions of 0.8\% and 1.4\%, respectively.
Roman supernovae do not directly tighten \(H_0\), but through parameter correlations they stabilize the inference and enable an absolute calibration of the supernova magnitude \(M_B\) that is competitive with, and potentially tighter than, current distance-ladder determinations.

For dark energy in the \(w_0w_a\)CDM framework, combining Roman supernovae with third-generation bright sirens substantially contracts the \(w_0\)–\(w_a\) contour, yielding a Figure of Merit that is competitive with, and for CE+ET+LVK+Roman exceeds, the latest DESI DR2 BAO plus DESY5 supernova combination. By contrast, \(\Omega_m\) benefits less from bright sirens because most events lie at low redshift, where the expansion is less sensitive to the matter density.

We also examined the impact of sky-localization thresholds on $H_0$ precision. For 3G networks, localization cuts of $\Delta\Omega<50~\mathrm{deg}^2$ cause only modest degradation, while $\Delta\Omega<10~\mathrm{deg}^2$ leads to more visible penalties. Nevertheless, subpercent precision is still reached within five years, dropping below 1\% for ET+LVK and 0.5\% for CE+ET+LVK. These results suggest that EM follow-up strategies can be optimized by prioritizing events localized within $\sim50~\mathrm{deg}^2$, or more aggressively $\sim10~\mathrm{deg}^2$, without significantly compromising cosmological precision.

Finally, we stress that the detector combinations considered here are forecast scenarios rather than firm predictions, since there is no guaranteed plan for overlap between the current 2G LVK facilities and future 3G observatories such as ET and CE. While 3G instruments will be far more sensitive than 2G detectors, sensitivity alone does not ensure robust sky localization and three-dimensional event reconstruction when the network has only one or two sites, which are key requirements for identifying bright sirens and enabling cosmological inference. In this sense, continued operation of 2G facilities alongside early 3G operations provides a conservative and scientifically valuable baseline. An intermediate route could be LIGO--India, expected in the early 2030s, although its schedule remains uncertain; overall, these considerations motivate treating 2G+3G overlap as a plausible assumption and highlight the importance of long-term strategic investments in gravitational-wave infrastructure.

These conclusions rest on Fisher forecasts for most events (with per-event MCMC for LVK), simplified time-scaling, and an idealized Target-of-Opportunity model; they also omit BNS tidal effects in waveform generation and detailed scheduling systematics. Incorporating more realistic follow-up operations, improved waveform physics and calibration, lensing-dispersion mitigation, and a fuller treatment of selection and redshift systematics will refine the projections. Even with these caveats, bright sirens observed with third-generation detectors, especially when combined with Roman-quality supernova samples, can deliver decisive measurements of \(H_0\) and competitive constraints on dynamical dark energy, establishing GW+EM cosmology as a cornerstone of next-decade precision cosmology.


\section*{Acknowledgments}
RM acknowledges financial support from FAPES (Brazil) and CAPES (Brazil) and thanks the Institute for Theoretical Physics at the University of Heidelberg for hosting his stay in Heidelberg.
VM acknowledges partial support from CNPq (Brazil) and FAPES (Brazil).
The authors would like to acknowledge the use of the computational resources provided by the \href{https://computacaocientifica.ufes.br/scicom}{Sci-Com Lab} of the Department of Physics at UFES, which was funded by FAPES, CAPES and CNPq.



\bibliographystyle{JHEP}
\bibliography{biblio.bib}

\end{document}